\documentstyle[12pt,epsf]{article}
\newcommand{\ee}{\end{equation}}
\newcommand{\be}{\begin{equation}}

\begin{document}
\title{Sphaleron-Bisphaleron bifurcations in a 
custodial-symmetric two-doublets model}

\author{ {\large Y. Brihaye}$^{\diamond}$
\\
$^{\diamond}${\small Physique-Math\'ematique,
Universit\'e de Mons-Hainaut}\\
{\small B-7000 Mons, Belgium}
\\
}
\date{\today}
\maketitle
\begin{abstract}
The standard electroweak model is extended by means of a second
Brout-Englert-Higgs-doublet. The symmetry breaking potential
is chosen is such a way that (i) the Lagrangian possesses a custodial
symmetry, (ii) a static, spherically symmetric
ansatz of the bosonic fields consistently 
reduces the Euler-Lagrange
equations to a set of differential equations.
The potential involves, in particular,
products of fields of the two doublets,
 with a coupling constant $\lambda_3$.
Static, finite energy solutions of the classical equations
are constructed. The regular, non-trivial solutions having
the lowest classical energy can be of two types~:
 sphaleron or bisphaleron, according to the coupling constants.
A special emphasis is put to the bifurcation
between these two types of solutions which is analyzed  in
function of the different constants of the model,
 namely of $\lambda_3$.
\end{abstract}
\medskip \medskip
\newpage
\section{Introduction}
It is known for a long time \cite{toof} that baryon and lepton
numbers are not strictly conserved in the standard model of
electroweak interactions (see \cite{ru}, \cite{trodden} for reviews).
Baryon number violating processes \cite{shap} 
involve the crossing through an
energy barrier separating topologically
inequivalent vacua of the underlying gauge theory.
Remarquably,  this energy barrier is high but finite. It corresponds
to a static, regular
solution of the classical equations of motion~: the sphaleron \cite{ma}.
The sphaleron was first constructed in the case $\theta_W = 0$
($\theta_W$ denotes the Weinberg angle) where
a consistent spherically symmetric ansatz \cite{km}
transforms the Euler-Lagrange equations of the theory
into differential equations.
 The  Klinkhamer-Manton (KM) sphaleron is, however, 
not the minimal energy barrier
when the mass of the Brout-Englert-Higgs-boson (BEH-boson)
exceeds some critical value. Indeed, for $M_H >> M_W$ \cite{bk,yaffe,kb}
another branch of solutions exists which bifurcate from the sphaleron
branch  for $M_H \approx 12 M_W$. The new solutions have a lower
energy than the sphaleron and since they appear by pairs connected
each other by the parity operator, they were called {\it bishpaleron}.
Nowadays the possibility that bisphalerons constitute the energy barrier
allowing for baryon number  violating process is ruled out in the
minimal (one doublet) electroweak model by the perturbative
upper limit of the BEH-field
and the sphaleron-bisphaleron  bifurcation remains
a curiosity of the classical equations.

However several extensions of the minimal Weinberg-Salam
model are curently under investigation as alternative
candidates for the description of electroweak interactions
(see e.g. \cite{sophie,hunter}).
Among these various extensions,
the ones incorporating more than one multiplet of BEH-boson play
a central role. For instance, the minimal supersymmetric electroweak model,
considered for many theoretical reasons, involves two
BEH-doublets.
These extended models lead generally to involved classical equations
where the generalisations of sphaleron and bisphaleron can be looked
for, as well as eventual other type of solutions of soliton type.
It particular, it is challenging to study the domain of parameters
for which bisphaleron exist and  to see if it intersect the domain
of physically acceptable parameters.
This question was adressed early in \cite{btt} and in \cite{kleihaus}.
The potential used  these papers does not involve a direct coupling
between the doublets. Here we will extend the potential choosen
in these two papers by mean of a supplementary interaction  between the
two BEH-doublets. The influence of the new term on the
sphaleron-bisphaleron bifuraction will be studied in details.

To be complete let us mention that the classical equations of the
two-doublets-extended standard model were also investigated in
\cite{peccei, hindmarsh} with even more general potential
but, to our knowledge, these authors did not put the emphasis
on  bifurcation between the two types of lowest energy solutions.

In Sect.2, we present the model, the notations and the
physical parameters.  The spherically symmetric
ansatz, the equations and boundary conditions are given in Sect.3;
the numerical solutions are then discussed in Sect.4.

\section{The model}
The lagrangian that we consider in this paper reads
\be
\label{lag}
{\cal L} = -\frac{1}{4} F_{\mu\nu}^a F^{\mu\nu,a}
+ (D_\mu \Phi_{(1)})^{\dagger} (D^\mu \Phi_{(1)})
+ (D_\mu \Phi_{(2)})^{\dagger} (D^\mu \Phi_{(2)})
- V(\Phi_{(1)}, \Phi_{(2)})
\ee
Where $\Phi_{(1)}, \Phi_{(2)}$ denote the two BEH-doublets
and the standard definitions are used  for the covariant derivative
and gauge field strengths~:
\be
\label{cova}
F_{\mu\nu}^a=\partial_\mu V_\nu^a-\partial_\nu V_\mu^a
            + g \epsilon^{abc} V_\mu^b V_\nu^c
\ee
\be
D_{\mu} \Phi = \Bigl(\partial_{\mu}
             -\frac{i}{2}g \tau^a V_{\mu}^a
               )\Phi
\ee
(the limit $\theta_W=0$, i.e. $g'=0$, is used throughout the paper).

The most general gauge invariant potential constructed with 
two BEH-doublets
is presented namely in \cite{hunter}, it depends on nine constants.
Here we consider  the family of potentials of the form
\be
\label{pot}
     V(\Phi_{(1)}, \Phi_{(2)})
    = \lambda_1 (\Phi_{(1)}^{\dagger} \Phi_{(1)} - {v_1^2 \over 2})^2
    + \lambda_2 (\Phi_{(2)}^{\dagger} \Phi_{(2)} - {v_2^2 \over 2})^2
    + \lambda_3  (\Phi_{(1)}^{\dagger} \Phi_{(1)} - {v_1^2 \over 2}) 
    (\Phi_{(2)}^{\dagger} \Phi_{(2)} - {v_2^2 \over 2})
\ee
depending on five parameters. 
The terms directly coupling the two doublets is parametrized
by the constant $\lambda_3$
and one of  the main characteristic about this  potential resides 
in the fact that
it imposes a symmetry breaking mecanism to each of the BEH-doublets.
The case $\lambda_3 = 0$ is studied
at length in \cite{btt,kleihaus}.

The lagrangian (\ref{lag}) is off course invariant under SU(2) gauge
transformations but it further possesses a larger global symmetry
under SU(2) $\times$ SU(2) $\times$ SU(2).
In fact, the part of the lagrangian (\ref{lag}) involving
the scalar fields  can be
written in terms of 2$\times$2 matrices defined by
\be
      M_{1,2} \equiv
       \left(\matrix {\phi_0^*& \phi_+\cr
                      -\phi_+^* & \phi_0\cr}\right)_{1,2}
                \ \ \ {\rm for} \ \ \
  \Phi_{1,2} =  \left(\matrix {\phi_+ \cr \phi_0\cr } \right)_{1,2}
\ee
When written  in terms of the matrices $M_1$ and $M_2$, the lagrangian
(\ref{lag}) becomes manifestly invariant under the transformation
\be
\label{custo}
   V_{\mu}' = A V_{\mu} A^{\dagger} \ \ , \ \
   M_1' = A M_1 B  \ \ , \ \
   M_2' = A M_2 C
\ee
 with $A,B,C \in$ SU(2); this the custodial symmetry.
The double symmetry breaking mechanism imposed by the potential
(\ref{pot}) leads to a mass $M_W$ for two of the three gauge
vector bosons and, namely, to two neutral BEH-particles
with masses $M_h$, $M_H$.
In terms of the parameters of the Lagrangian, these masses are
given by  \cite{hunter,herquet}
\be
\label{masse}
     M_W = {g \over 2} \sqrt{v_1^2 + v_2^2} \ \ \ , \ \ \
M^2_{H,h} = \frac{1}{2}[A_1 + A_2 \pm \sqrt{(A_1-A_2)^2 + 4 B^2}]
\ee
with
\be
    A_1 = 2 v_1^2(\lambda_1 + \lambda_3) \ \ , \ \
    A_2 = 2 v_2^2(\lambda_2 + \lambda_3) \ \ , \ \
    B = 2 \lambda_3 v_1 v_2
\ee
For later convenience we also define
\be
\label{param}
    \tan \beta = {v_2 \over v_1} \ \ , \ \ 
    \rho_{H,h} = {M_{H,h} \over M_W}  \ \ , \ \
\epsilon_p = 4 {\lambda_p \over g^2} \ \ , \ \ p=1,2,3
\ee
Note that the mass ratio $\rho_{1,2}$ used in \cite{kleihaus}
are related to $\rho_{H,h}$ by 
$\rho_H = {\rm max}\{\rho_1,\rho_2\}$,
$\rho_h = {\rm min}\{\rho_1,\rho_2\}$.
For physical reasons,  we consider only $v_1 \geq 0$, $v_2 \geq 0$
so that $0 \leq \beta \leq \pi/2$. Interestingly, the parameter 
$\epsilon_3$ can be negative but cannot take arbitrary values. 
The following relations are usefull to determine the physical region~:
\begin{eqnarray}
\epsilon_1 \cos^2 \beta + \epsilon_2 \sin^2 \beta 
&=& \frac{1}{2}\rho_H^2 + \rho_h^2 -  \epsilon_3 \nonumber \\
\epsilon_1 \cos^2 \beta - \epsilon_2 \sin^2 \beta 
 &=& \frac{1}{2} \sqrt{(\rho_H^2 - \rho_h^2)^2 - 4 \epsilon_3^2 \sin^2(2 \beta)}
- \epsilon_3 \cos(2 \beta)
\end{eqnarray}
The physical domain is then determined by the conditions
\begin{equation}
\frac{\rho_h^2-\rho_H^2}{2 \sin(2\beta)} \leq \epsilon_3   \leq
\frac{\rho_H^2-\rho_h^2}{2 \sin(2 \beta)} \ \ , \ \
              0 \leq  \rho_h^2 \leq \rho_H^2
\end{equation}
Physical domains are presented on Fig.1 for the case $\rho_h=3$
for $\beta = \pi / 4$ and $\beta = 3 \pi /8$.

\section{Spherical symmetry}
In order to construct classical solutions of the Lagrangian (\ref{lag}),
we perform a spherically symmetric ansatz for the fields.
With the notations of \cite{akiba}, it reads
\begin{eqnarray}
\label{sphsym}
V^a_0 &=& 0\nonumber\\
V^a_i &=& 
{1-f_A(r)\over{gr}} \epsilon_{aij}\hat r_j 
+{f_B(r)\over{gr}}(\delta_{ia}-\hat r_i \hat r_a)
+{f_C(r)\over{gr}}\hat r_i \hat r_a\nonumber\\
\phi_{(1)} &=& {v_1\over{\sqrt{2}}} [H(r) + i (\hat r.\vec \sigma)K(r)]
\left(\begin{array}{c} 0\\ 1\end{array}\right)\nonumber\\
\phi_{(2)} &=& {v_1\over{\sqrt{2}}} [\tilde H(r) + i (\hat r.\vec \sigma)\tilde K(r)]
\left(\begin{array}{c} 0\\ 1\end{array}\right)
\end{eqnarray}
Where $f_A, f_B, F_C, H, K, \tilde H, \tilde K$ are real radial
functions. It can be shown that the above ansatz transforms
the Euler-Lagrange equations into a set of coupled differential 
equations.
The custodial symmetry has been used to set the two doublets
parallel to each other
asymptotically. The condition $V_0 = 0$ results from a gauge fixing.
In fact, the spherically symmetric ansatz
leaves a residual gauge symmetry which can be exploited
to eliminate one of the seven
radial functions \cite{bk,yaffe,akiba}.
Here we will adopt the radial gauge
$x_j V^a_j = 0$ which implies $f_C = 0$.

The effective one-dimensional energy density can be obtained 
after some algebra
\begin{eqnarray}
   E&=&{M_W \over \alpha_W} \int_0^{\infty} {\cal E} dx \nonumber \\
    &\equiv &{M_W \over \alpha_W} \tilde E \
\end{eqnarray}
with $\alpha_W \equiv {g^2 \over 4 \pi}$ and
\begin{eqnarray}
\label{enden}
{\cal E} &=&
(f'_A)^2+(f'_B)^2+{1\over{2x^2}}  (f^2_A+f^2_B-1)^2
\nonumber  \\
&+& \cos^2\beta [(H(f_A-1)+Kf_B)^2+(K(f_A+1)-Hf_B)^2+2x^2((H')^2+(K')^2)
\nonumber\\
&+& (\tilde H (f_A-1)+ \tilde K f_B)^2+(\tilde K (f_A+1)-\tilde Hf_B)^2
+ 2x^2((\tilde H')^2+(\tilde K')^2) ]\nonumber\\
&+& \cos^4\beta [\epsilon_1 x^2(H^2+K^2-1)^2
+\epsilon_2 x^2(\tilde H^2 + \tilde K^2-tg^2\beta)^2 \nonumber \\
 &+& \epsilon_3 x^2 (H^2+K^2-1)(\tilde H^2+ \tilde K^2-tg^2\beta)]
\end{eqnarray}
The dimensionless variable $x=M_W r$ is used and 
the prime means derivative with respect to $x$.
The equations to solve can then be obtained by varying the functional
(\ref{enden}) with respect to the six radial functions.
Remark that in the case
$v_2=0,\lambda_2=\lambda_3=0$ the equations for $\tilde H , \tilde K$ decouple
and these functions can be set consistently to zero;
the remaning system just corresponds to the one of the 1DSM.

The conditions of regularity of the solutions at the origin
imposes namely $f_A^2 + f_B^2 = 1$ at $x=0$,
the custodial symmetry (\ref{custo}) can then be exploited to
fix the following values of the radial fields at the origin
\be
\label{condin}
  f_A(0) = 1 \ \ , \ \ f_B(0) = 0 \ \ , \ \
  K(0) = 0 \ \ , \ \ \tilde K(0) = 0 \ \ , \ \
  H'(0) = 0 \ \ , \ \ \tilde H'(0) = 0
\ee
On the other hand,
the condition of finiteness of the classical energy imposes
the following asymptotic form
\begin{eqnarray}
\label{condas}
&(f_A,f_B)_{x=\infty} &= (\cos 2\pi q , \sin 2 \pi q) \nonumber \\
&(H,K)_{x=\infty} &= (\cos \pi(q-k),\sin \pi(q-k)) \nonumber \\
&(\tilde H, \tilde K)_{x=\infty} &= \tan \beta \ (\cos \pi q,\sin \pi q)
\end{eqnarray}
for some real number $q$ and  $k$ equal to zero or one.
For later use we define $q\equiv 1/2 + \delta$.

\section{Discussion of the solutions}
In order to make the following discussion self contained, 
we first summarize the
main features of the solutions available in 
the one doublet standard model (1DSM), i.e. in the case
$v_2=\lambda_2=\lambda_3=0$ with  $\tilde H =  \tilde K = 0$.
\subsection{1DSM}
There exist at least one solution, the Klinkhamer-Manton (KM) sphaleron, 
for all values of  $\rho_1$ \cite{km} (
$\rho_H \equiv \rho_1 $ in this case).
For this solution one can further set $f_B = H = 0$ by an appropriate
choice of the custodial symmetry;
the classical energy increases monotonically
as a function of $\rho_1$~:
\be
   \tilde E_s(\rho_1=0) \approx 3.04 \ \ ,
   \ \ \tilde E_s(\rho_1=\infty) \approx 5.41
\ee
The KM sphaleron is always unstable but the number of its
directions of instability increase
when $\rho$ increases \cite{yaffe,bksta}.
At $\rho \approx 12.04$ a couple of new solutions, the bisphalerons,
bifurcate from the sphaleron. The two bisphalerons
 (which transform into each other by
parity) have the same energy and their energy is lower than the one
of the KM sphaleron
\be
   \tilde E_{bs}(\rho_1=12.04) = \tilde E_s(\rho_1=12.04) \approx 4.86 \ \ , \ \
   \tilde E_{bs}(\rho_1=\infty) \approx 5.07
\ee
The parameter $q$ defined in (\ref{condas}) 
is equal to $1/2$ for sphaleron.
For bisphaleron, the deviation from  $q=1/2$ parametrized by $\delta$
varies from zero (at the bifurcation point)
  and  $\delta = \pm 0.06 $
  (at $\rho_1 = \infty$),   respectively  for the two bisphalerons.

\subsection{2DSM, case $\epsilon_3 = 0$}
Solutions of the Lagrangian (\ref{lag}) with $\lambda_3=0$
were first constructed  in \cite{btt} and  reconsidered in \cite{kleihaus}
where the emphasis on the sphaleron-bisphaleron bifurcation was set.
As in 1DSM, sphaleron have
          $f_B = H = \tilde H = 0$ and  seem to
exist for all values of the parameters of the potential.
The angle parameter $q$ defined in (\ref{condin}) correponds to $q=1/2$,
irrespectively of the coupling constants of the potential.

By contrast, the six radial functions corresponding to the bisphaleron
are non trivial and fullfil
the boundary conditions (\ref{condas}).
The parameter  $q$ depends of the various coupling constant  
although remaining
 close to $1/2$
(e.g. $\delta =0.036$ for $\rho_1= 14, \rho_2 = 1, \beta = 0.2$).

The results of \cite{kleihaus} show the existence  of a smooth
surface in the $\rho_1, \rho_2, \beta$-parameter space inside of which
only sphaleron solution exist while sphaleron and bisphaleron coexist
outside, the bifurcation taking place {\it on} the surface.
The critical surface can be determined only numerically
by studying a few parameters characterizing 
the bisphaleron solutions, namely
the values $\delta, K(0), \tilde K(0)$, as functions 
of $\beta, \epsilon_{1,2,3}$.
Varying one of these parameters and fixing the other three, 
a critical point
is determined when $\delta, K(0), \tilde K(0)$ approach zero. 
This is done on
Fig. 2 for $\beta = \pi/4$ , $\epsilon_3=0$ and $\rho_H = 3 \rho_h$. 
The critical
point then corresponds to $\rho_H \approx 2.397$.  
It is worth noticing that bisphaleron solutions with $\delta = 0$ 
but $K(x) \neq 0$, $\tilde K(x) \neq 0$
also occur outside the critical surface. 
This is illustrated on Fig.3 where we set
$\rho_H^2 + \rho_h^2 = 9$ and vary $\Theta \equiv atan(\rho_H/\rho_h)$. 
Clearly,
$\delta$ goes to zero in the limit $\Theta \rightarrow \pi/4$.
The figures 2 and 3 are complementary to the ones presented in 
\cite{btt, kleihaus}

The main feature of the bisphaleron solutions in the 2DSM is that the angle
$\tilde \phi = \arctan( \tilde K / \tilde H)$
increases monotonically from
$0$ (for $x=0$) to $ \pi- q$ (for $x= \infty$) while
$\phi = \arctan( K / H)$ decreases from $0$ to $\pi(q-1)$;
that is to say that these solutions of lowest energy have  $k=1$ in (\ref{condas}).
In fact, non trivial solutions with $k=0$ were discovered in \cite{btt} but
since they have a higher energy, they are likeky less interesting
as far as the energy barrier is concerned.

\subsection{2DSM, case $\epsilon_3 \neq 0$}
The construction of solutions in the domain $\rho_h, \rho_H,
\beta, \epsilon_3$ and the study of the critical hypersurface
in this four-parameter space is a vast task. For definiteness,
we have limited our investigations to the case $\rho_h = 3.0$ 
and to two values of $\beta$, namely  $\beta = \pi/4$ and 
$\beta = 3 \pi / 8$. 

We first discuss the results for $\beta = \pi/4$, i.e. $v1 = v_2$.
Here are a few "points`` on the bifurcation
line correponding to $\beta = \pi/4$~:
\begin{equation}
(\rho_h , \rho_H) : (10, 7.6) \ , \ (3.0, 7.08) \ , \
                    (5.0, 6.08) \ , \ (5.585, 5.585)
\end{equation}
Our main concern is to determine how this domain
of the $\rho_h, \rho_H$ plane evolves with $\epsilon_3$.
The result is illustrated by Fig. 1.  
where the physical domain is delimited by the solid lines.
 The sphaleron-bisphaleron bifurcation line
is represented by the solid line with bullets and the domains where
sphaleron only exist is indicated, as well as the domain where
sphaleron and bisphaleron coexist.
The critical line separating these two regions
clearly exhibits two different behaviours
on the domain of parameters considered~:
for $ 6.2 \leq \rho_H \leq 8$, the critical line is roughly
a function increasing linearly with the coupling constant
$\epsilon_3$. As a consequence, the critical line 
intersect the lower line delimiting the physical domain,
for instance at $\rho_H \approx 6.22 \ ,\ \epsilon_3 \approx -15.0$.
Clearly, for the negative values of $\epsilon_3$,
the the minimal mass of $M_H$ (with all other parameters
fixed) for which bisphaleron exist is lowered by the presence
of a direct interaction between the BEH-fields.
For $8.0 < \rho_H < 10.0$  the critical lines develops
a plateau at $\epsilon_3 \approx 10.0$
and $\rho_{H,cr}$ depends only weakly of $\epsilon_3$.
For some unknown reason, the numerical
analysis become very difficult when the critical line appraoches 
the limit of the physical domain.

The behavior of the critical line turns out to be completely different
for $\beta = 2 \pi/8$, in this case, the limit of the physical domain 
is indicated by the dashed lines and the critical line bu the dashed line
with the triangles.  
In contrast to the case $\beta = \pi/4$ we see here
that the critical value 
$\rho_{H,cr}$ decreases roughly linearly with $\epsilon_3$.
For $\epsilon_3 > 0$ the barrier turns out to be a bisphaleron 
for lower values
of $\rho_H$ than in the $\epsilon_3 = 0$ case.  
Here, the critical line seems to cross the two lines determining 
the physical domain
where it naturally terminate.

We further studied the critical phenomenon 
for $3 \pi / 8 < \beta < \pi / 4$
and observed a smooth evolution of the 
critical lines displayed in Fig. 1.

\section{Conclusion}
The lagrangian considered in this note leads to a tricky
system of six differential equations with boundary conditions
and depending effectively on four parameters. 
Many types of non-trivial solutions
 can be constructed numerically \cite{btt} but,
at the moment, the ones with lowest
energy are identified as the sphaleron or the bisphaleron, depending
on the different coupling constants.
The determination
of the critical hypersurface of bifurcation
in the space of parameters constitutes a huge task
which can be studied only numerically. 
The problem is for a large part academical;
however at the moment, the theoretical limits on the BEH-boson 
masses
obtained in  the two-doublets extension of the electroweak model,
do not yet exclude that the energy barrier between
topologically different vacua could be determined by a bisphaleron.
A few years ago, it was already pointed out that the minimal
mass of the neutral BEH-fields
for the barrier to be of the bisphaleron type
is considerably lower in the two-doublets model (without
direct interaction of the doublets in the potential)
than in the minimal, one-doublet model. The calculations reported here
suggest that, if a custodially-invariant coupling term is
supplemented to the potential,  the critical mass $\rho_{H,cr}$
varies roughly linearly with the new coupling constant $\epsilon_3$.
The supplementary coupling constant cannot off course be
arbitrarily large, but on the physical domain, the critical
value $\rho_{H,cr}$ indeed  dimnishes, all other masses being fixed.
Off course, many other terms can be added in the potential
\cite{hunter} and it could be that bisphaleron solution exist for
still lower masses of the neutral BEH-field when terms
allowing for charged BEH-fields are considered as well.

\bigskip
\bigskip
{\bf Acknowledgements}
This work started from several discussions with Michel Herquet.
I gratefully acknowledge him for these discussions, his interest 
in the topic and for numerous private communications about his own
reseach \cite{herquet}.

\newpage
\small{

 }

\newpage
\begin{figure}
\centering
\epsfysize=20cm
\mbox{\epsffile{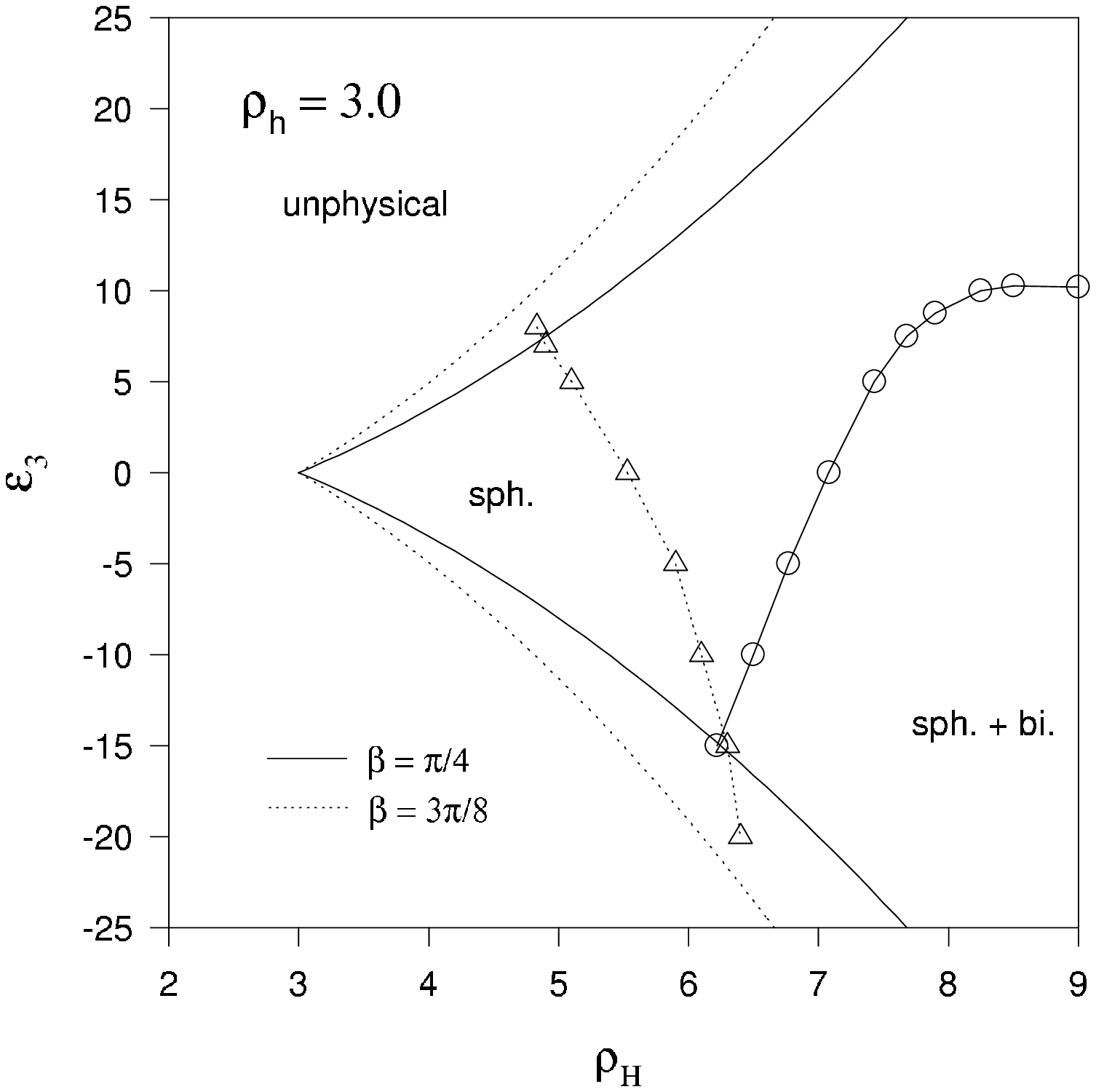}}
\caption{\label{Fig.1} The physical domain corresponding to the
case $\rho_h =3.0$, $\beta = \pi/4$ (resp. $\beta = 3 \pi/8$)  
are represented  by the solid (resp. dashed) line in the
$\rho_H , \epsilon_3$
plane. The line with bullets (resp. triangle) represents
the sphaleron-bisphaleron bifurcation. }
\end{figure}
 \newpage
\begin{figure}
\centering
\epsfysize=20cm
\mbox{\epsffile{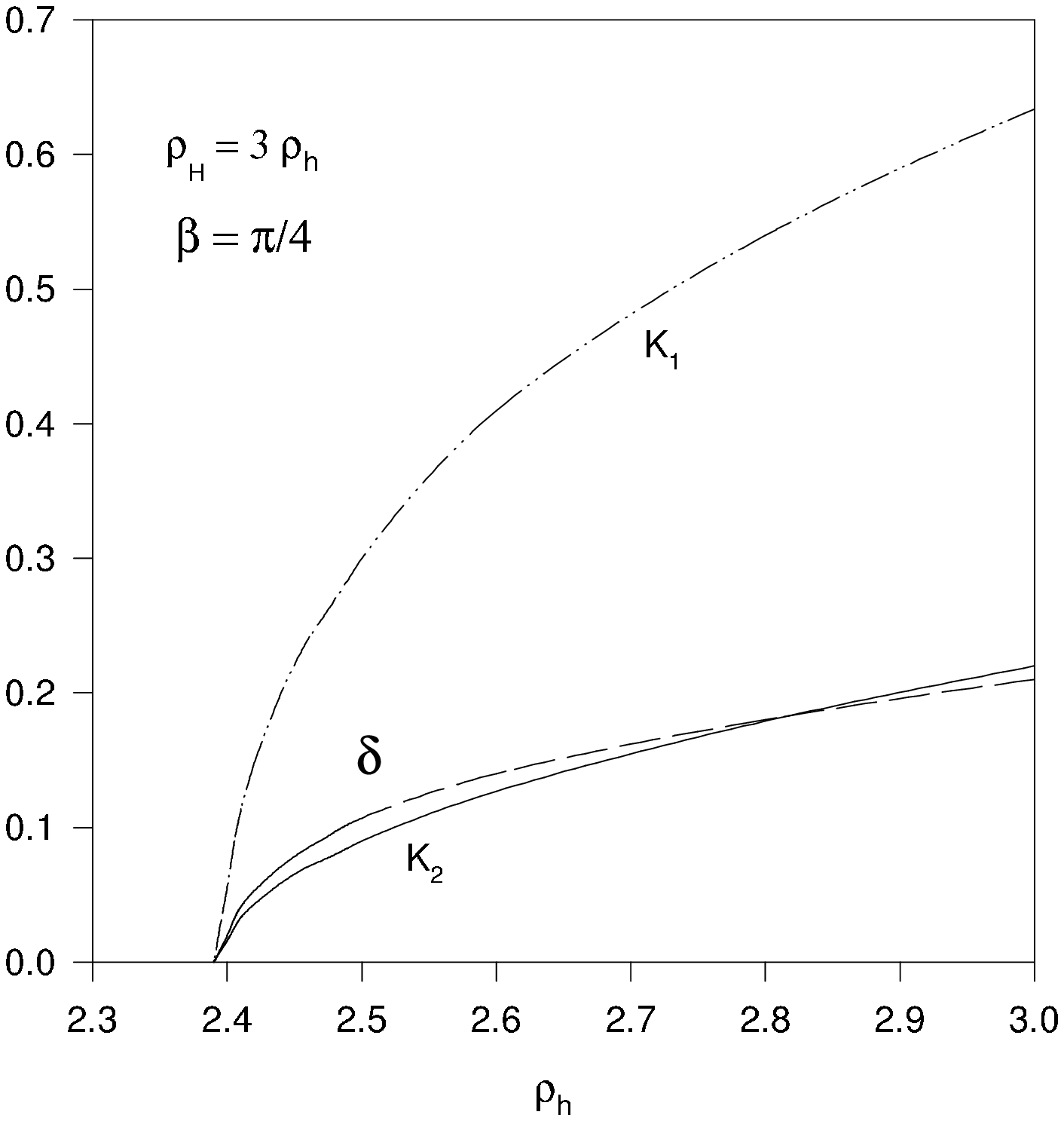}}
\caption{\label{Fig.2}The evolution of the parameters 
$\delta, K_1 \equiv K(0), K_2 \equiv \tilde K(0)$
is shown as function of $\rho_h$ for $\rho_H = 3 \rho_h$.
}
\end{figure}
 \newpage
\begin{figure}
\centering
\epsfysize=20cm
\mbox{\epsffile{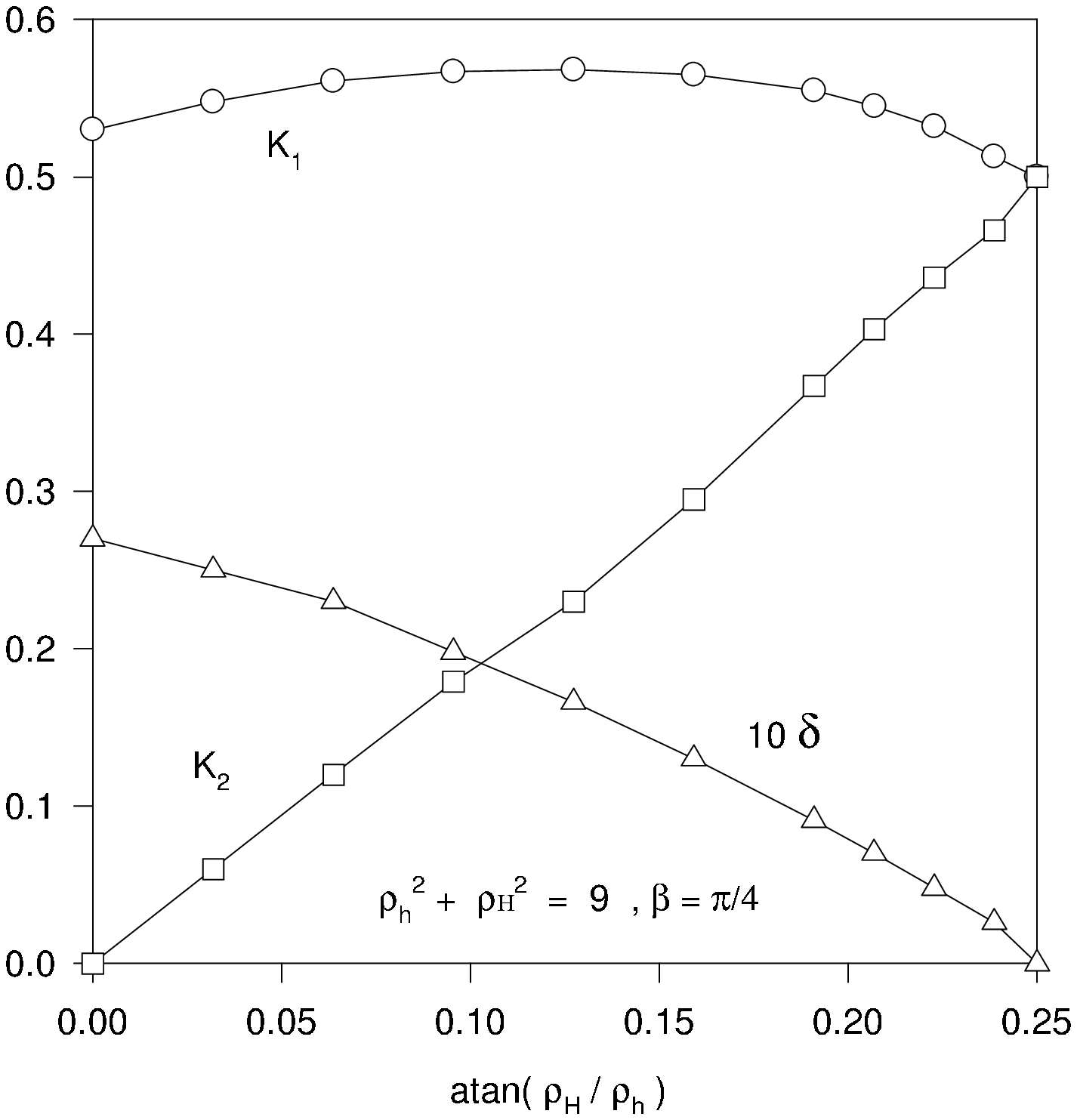}}
\caption{\label{Fig.3}The evolution of the parameters
$\delta, K_1 \equiv K(0), K_2 \equiv \tilde K(0)$
is shown as function 
of $\rho_H / \rho_h$ for $\rho_H^2 + \rho_h^2  =9$. }
\end{figure}
 \newpage


\begin{thebibliography}{99}
\bibitem{toof}
't Hooft, G.,
Symmetry breaking through Bell-Jackiw Anomalies,                             
{\bf Phys. Rev. Lett.}37, 8  (1976).
\bibitem{ru}
Rubakov, V.A., Shaposhnikov, M.~E.,
Electroweak baryon number non-conservation in the early universe and
in high energy collisions,
{\bf Phys. Usp.}39, 461 (1996), hep-ph/9603208.
\bibitem{trodden}
Trodden, M.,
Electroweak baryogenesis,
CWRU-P6-98 , hep-ph/9803479.
\bibitem{shap}
Kuzmin, V.~A., Rubakov, V.~A., and Shaposhnikov, M.~E.,
On anomalous electroweak baryon-number non-conservation                      
in the early universe,                                                       
{\bf Phys. Lett. B}155, 36 (1985).                                                   
\bibitem{ma} 
Manton, N.~S.,                                                           
Topology in the Weinberg-Salam theory,                                       
{\bf Phys. Rev. D}28, 2019 (1983).                                                   

\bibitem{km}
Klinkhamer, F.~R., and Manton, N.~S.,                                   
A saddle-point solution in the Weinberg-Salam theory,                        
{\bf Phys. Rev. D}30, 2212  (1984).                                                   


\bibitem{bk}
Kunz, J., and Brihaye, Y., (1989),                                         
New sphalerons in the Weinberg-Salam theory,                                 
{\bf Phys. Lett. B}216, 353 (1989).                                            

\bibitem{yaffe}
Yaffe, L.~G.,                                                            
Static solutions of SU(2)-Higgs theory,                                      
{\bf Phys. Rev. D}40, 3463 (1989).                                                   

\bibitem{kb}
Brihaye, Y., and Kunz, J.,                                               
A sequence of new classical solutions                                        
in the Weinberg-Salam model,                                                 
{\bf Mod. Phys. Lett. A}4, 2723 (1989).                                      

\bibitem{sophie}
Monig, K., Limits on $m_H$ Present and Future,
Delphi collaboration report 98-14 PHYS 764, (1998).  

\bibitem{hunter}
Dawson, S, Gunion, J.F., Haber, H.E. and Kane, G.L.,
 The Higgs Hunters'guide,
Frontiers in Physics, Addison-Wesley (1990).
 
\bibitem{btt}
Bachas, C., Tinyakov, P. and Tomaras, T.N., 
On spherically-symmetric solutions in the two-Higgs standard
model,
{\bf Phys. Lett. B}385, 237 (1996).

\bibitem{kleihaus}
 Kleihaus, B.,
Energy barrier in the two-Higgs model,
{\bf Mod. Phys. Lett. A} 14, 1431 (1999).

\bibitem{peccei} Kastening, B.,  Peccei, R.  and  Zang, X.,
Sphaleron in the two doublet Higgs model,
{\bf Phys. Lett. B} 266, 413 (1991).


\bibitem{hindmarsh} Grant, J. and  Hindmarsh M.,
Sphaleron in Two Higgs Doublets Theories,
{\bf Phys. Rev. D} 64, 016002 (2001).

\bibitem{herquet} Herquet, M., Sym\'etries custodiales
dans les mod\`eles \`a deux doublets de Higgs, DEA Thesis,
Universit\'e Catholique de Louvain (Sept. 2004).

\bibitem{akiba}
Akiba, T., Kikuchi H., and Yanagida, T.,
The free energy of the sphaleron in the Weinberg-Salam model,
{\bf Phys. Rev. D}40, 179 (1989).                                              


\bibitem{bksta}
Brihaye, Y., and Kunz, J.,                                               
Normal modes around the SU(2) sphalerons,                              
{\bf Phys. Lett. B}249, 90 (1990).                                          
\end{thebibliography}
\end{document}